\documentclass[twocolumn,showpacs,preprintnumbers,amsmath,amssymb]{revtex4}

\usepackage{graphicx}
\usepackage{dcolumn}
\usepackage{bm}

\begin{document}


\title{Anomalous Transport through the $p$-Wave Superconducting Channel in the 3-K Phase of Sr$_2$RuO$_4$}

\author{H. Kambara$^1$, S. Kashiwaya$^1$, H. Yaguchi$^2$, Y. Asano$^3$, Y. Tanaka$^4$, and Y. Maeno$^5$} 
\affiliation{
$^1$National Institute of Advanced Industrial Science and Technology (AIST), Tsukuba 305-8568, Japan\\
$^2$Department of Physics, Faculty of Science and Technology, Tokyo University of Science,
Noda 278-8510, Japan\\
$^3$Department of Applied Physics, Hokkaido University, Sapporo 060-8628, Japan\\
$^4$Department of Applied Physics, Nagoya University, Nagoya 464-8603, Japan\\
$^5$Department of Physics, Kyoto University, Kyoto 606-8502, Japan}


\date{\today}

\begin{abstract}
Using micro fabrication techniques, we extracted individual channels of 3-Kelvin (3-K) phase
superconductivity in Sr$_2$RuO$_4$-Ru eutectic systems and confirmed odd-parity superconductivity
in the 3-K phase, similar to pure Sr$_2$RuO$_4$.
Unusual hysteresis in the differential resistance-current and voltage-current characteristics
observed below 2 K indicates the internal degrees of freedom of the superconducting state.
A possible origin of the hysteresis is current-induced chiral-domain-wall motion due to the
chiral $p$-wave state.
\end{abstract}


\maketitle

Most superconductors have a spin-singlet pairing state, including high-$T_c$ cuprates.
Spin-triplet pairing superconductors are quite rare and the superconducting transition temperature
($T_c$) is generally low ($\sim 1$ K).
Layered perovskite Sr$_2$RuO$_4$ (SRO) is one of the best candidates for spin-triplet
pairing with a $T_c$ of 1.5 K, and the superconducting vector order parameter is similar to that
of the superfluid $^3$He-A---the so-called chiral $p$-wave state \cite{mackenzie}.
The pure SRO phase (1.5-K phase) has been well-studied, but SRO-Ru eutectics in which
Ru lamellae are embedded are not well understood.
The SRO-Ru eutectic is called the 3-Kelvin (3-K) phase \cite{maeno,ando,yaguchi} because of the
remarkable enhancements of $T_c$ up to 3 K.
However, the enhancement mechanism of $T_c$ and the pairing symmetry of the 3-K phase have not
been understood clearly.
Since the 3-K phase is the interface superconductivity in the SRO region between SRO and
Ru \cite{yaguchi}, the volume fraction of the superconducting state is very low compared with
that of pure phase.
Thus, it is difficult to determine the pairing symmetry by the ordinary method, i.e.
Knight shift by NMR.
It is extremely important to determine the pairing symmetry of the 3-K phase, because
$T_c$ of 3 K is the highest among spin-triplet superconductors if the 3-K phase is
established as a spin-triplet superconductor.
Moreover, the 3-K phase enables us to reveal nanoscale physics in inhomogeneous spin-triplet
superconductivity.

Thus far, several experiments have been performed to investigate the pairing symmetry of
the 3-K phase. In tunnel junction experiments, Mao {\it et al.} \cite{mao2} and
Kawamura {\it et al.} \cite{kawamura} observed the zero-bias conductance peak due to Andreev
resonance reflecting {\it non-s}-wave superconductivity \cite{tanaka}.
Hooper {\it et al.} \cite{hooper} reported the $c$-axis transport characteristics, which is
interpreted in terms of a complex Josephson network with anomalous asymmetric features in
their current-voltage characteristics.
Although these results imply an internal phase of superconducting order parameter, they do not
necessarily indicate odd-parity superconductivity of the 3-K phase.

\begin{figure}[ht]
\includegraphics[width=0.85\linewidth]{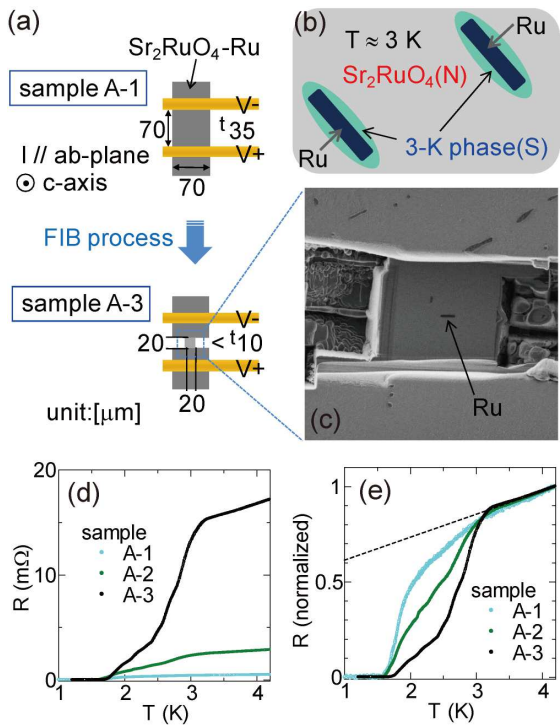}
\caption{\label{fig1}
(Color online)
Sketch of the sample configurations for samples A-1 ((a) top) and A-3 ((a) bottom) of
Sr$_2$RuO$_4$-Ru.
Sample A-3 was milled by FIB.
(b) Schematic image of nucleation of the 3-K phase superconductivity around Ru inclusions
at 3 K.
(c) Scanning ion microscope image ($50 \times 50$ $\mu$m$^2$) of sample A-3.
The black bars show the Ru inclusions.
(d) Zero-bias differential resistance ($R$) vs temperature for samples A-1, A-2, and A-3.
(e) Normalized resistance at 4.2 K vs $T$.
The dashed line represents a normal component.}
\end{figure}

In this study, we investigated the $ab$-plane differential resistance-current ($dV/dI-I$) and
voltage-current ($V-I$) characteristics of the 3-K phase by controlling the number of Ru
inclusions by a micro fabrication technique using a focused ion beam (FIB).
We extracted individual channels which connect the 3-K phase region/normal state region of
the SRO/3-K phase region as the S/N/S junction at 3 K $> T >$ 1.5 K (Fig. \ref{fig1}(b)).
We then determine the pairing symmetry of the 3-K phase from the temperature dependence of
the critical current ($I_c$).
Finally, we show quite unusual hysteresis in $dV/dI-I$ and $V-I$ below 2 K, which strongly
indicates the internal degrees of freedom, possibly due to the chiral $p$-wave state.

Eutectic crystals of SRO-Ru were grown in an infrared image furnace
by the floating zone method \cite{mao}.
The transport was measured using a standard four-probe technique.
The sample neck between the voltage-lead contacts was milled by FIB to reduce the number of Ru
inclusions in this region. 
The details of the sample preparation and measurement system are
described elsewhere \cite{kambara}.
For sample A, the neck region was $70 \times 70 \times 35$ $\mu$m$^3$ (sample A-1)
before the FIB process (Fig. \ref{fig1}(a) top).
Next, it was successively milled narrower and thinner (samples A-2$\rightarrow$A-3).
The final dimensions of the neck were $20 \times 20 \times (< 10)$ $\mu$m$^3$ (sample A-3)
(Fig. \ref{fig1}(a) bottom).
In Fig. \ref{fig1}(c), $\sim1$ $\mu$m thick and 1-6 $\mu$m long Ru inclusions are visible as black bars.
Only two pieces of Ru inclusions appear on the topmost surface in the neck region of sample A-3.
Thus, there should be only a few Ru inclusions in the neck region, including a few hidden below the surface.
Figure \ref{fig1}(d) shows the differential resistance at zero-bias current ($R$)-temperature curves of
samples A-1, 2, and 3, respectively.
For clarity, $R$ is normalized at 4.2 K, as shown in Fig. \ref{fig1}(e). 
We can see that the component of the 3-K phase is more dominant in sample A-3 than in sample A-1.
Thus, the milling process can change the ratio of the 1.5-K:3-K phase.
Note that it is not always the case that the component of the 3-K phase becomes dominant by milling.

\begin{figure}[ht]
\includegraphics[width=1.0\linewidth]{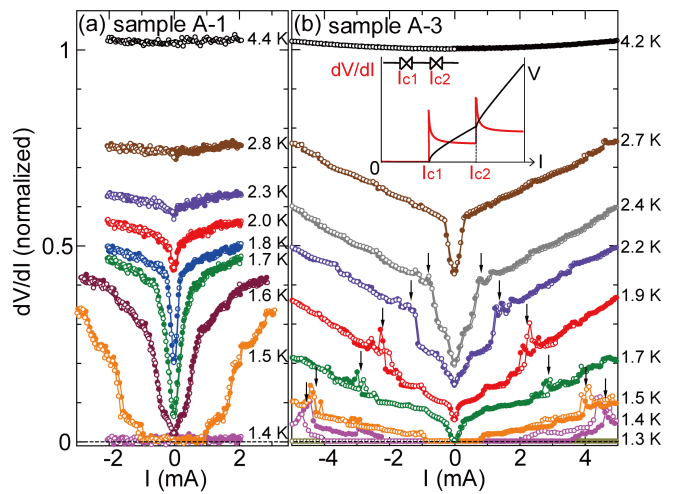}
\caption{\label{fig2}
(Color online)
Normalized $dV/dI$ vs $I$ as a function of temperature for samples A-1 (a)
and A-3 (b), respectively.
The open and filled symbols denote the different sweep directions, from zero to max. (open),
max. to min. (filled), and min. to zero (open).
A series of the most pronounced kinks are denoted by arrows (average currents of the $dV/dI$ peaks
in upward and downward sweep directions).
Inset: Characteristic kinks in the $dV/dI-I$ curves are explained by superconducting linkage channels
with critical currents ($I_{c1} < I_{c2} < ..$) in series. The red (black) line corresponds to
a schematic of $dV/dI-I$ ($V-I$).}
\end{figure}

In Fig. \ref{fig2}(a)(b), we show $dV/dI-I$ curves normalized to 4.2 K for samples A-1 and A-3.
Below $\sim 3$ K, $dV/dI$ curves show dip structures ($dV/dI \rightarrow 0$) near the zero-bias
current, reflecting the 3-K phase superconductivity.
Below 1.6 K, the $dV/dI$ curves show the flat zero resistance, reflecting the 1.5-K phase superconductivity
in addition to a path-formation due to the proximity effect of the 3-K phase superconductivity.
Increasing the bias currents beyond some critical values makes the $dV/dI$ values larger.
In particular, for sample A-3, more characteristic kinks were observed in the $dV/dI$ curves than
for sample A-1.
With a filament (non-uniform) model in which local superconducting channels connect Ru islands,
the characteristic kinks are naturally explained by the superconducting linkage
channels in series with their own critical currents, as illustrated in the inset of Fig. \ref{fig2}(b).
Thus we can extract the $I_c$ of each channel for sample A-3.
In sample A, it is difficult to separate the channels without FIB milling possibly because
similar linkage channels are averaged, i.e. sample A-1 hides the individuality of the channels.

\begin{figure}[ht]
\includegraphics[width=0.75\linewidth]{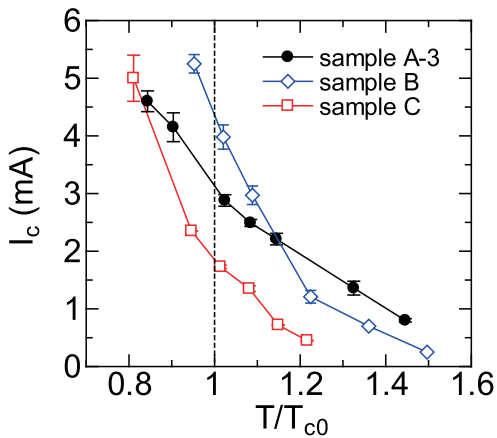}
\caption{\label{fig3}
(Color online)
Critical currents ($I_c$) vs normalized temperature for samples A-3, B, and C, respectively.
The three curves are traces of the most pronounced kinks in each series of $dV/dI-I$ curves
(For sample A-3, the arrows in Fig. \ref{fig2}(b)).}
\end{figure}

Figure \ref{fig3} shows the temperature dependence of $I_c$ focusing only on the most pronounced
kinks (the arrows in Fig. \ref{fig2}(b)) for three different types of samples cut from the same
crystal rod.
A continuous monotonic increase is seen in the value of $I_c$ with the temperature decreasing
through $T_{c0}$ which is defined as the zero-resistivity point in Fig. \ref{fig1}(d) and (e).
The temperature dependence of $I_c$ should reflect the relationship of the pairing symmetry of
the 3-K and 1.5-K phases.
From the experiment by Jin and co-workers \cite{jin}, $I_c$ of Josephson junctions (JJs) in Pb-SRO-Pb
increases below 7.2 K ($T_c$ of Pb) with decreasing temperature, but decreases below 1.3 K ($T_c$ of SRO).
This result is explained by the transition from $0$ to $\pi$-junction, which forms below 1.3 K
due to the difference in parity of the pairing symmetry (Pb has $s$-wave and SRO has
$p$-wave pairing symmetry) either as a first-order \cite{honerkamp} or a second-order
process \cite{yamashiro}.
Assuming a similar relation, we deduce that the 3-K and 1.5-K phases have the same parity,
such as $s/s/s$ or $p/p/p$ in the 3-K phase/1.5-K phase/3-K phase configuration, excluding
$\pi$-junction configurations.
Therefore, considering an odd-parity in the 1.5-K phase, the 3-K phase also has an odd-parity
in the local superconducting channel.
This is an experimental proof of the assumption in a phenomenological theory by Sigrist and Monien
(SM) \cite{sigrist}, which states that the filamentary phase at 3 K has $p$-wave pairing symmetry.

\begin{figure}[ht]
\includegraphics[width=1.0\linewidth]{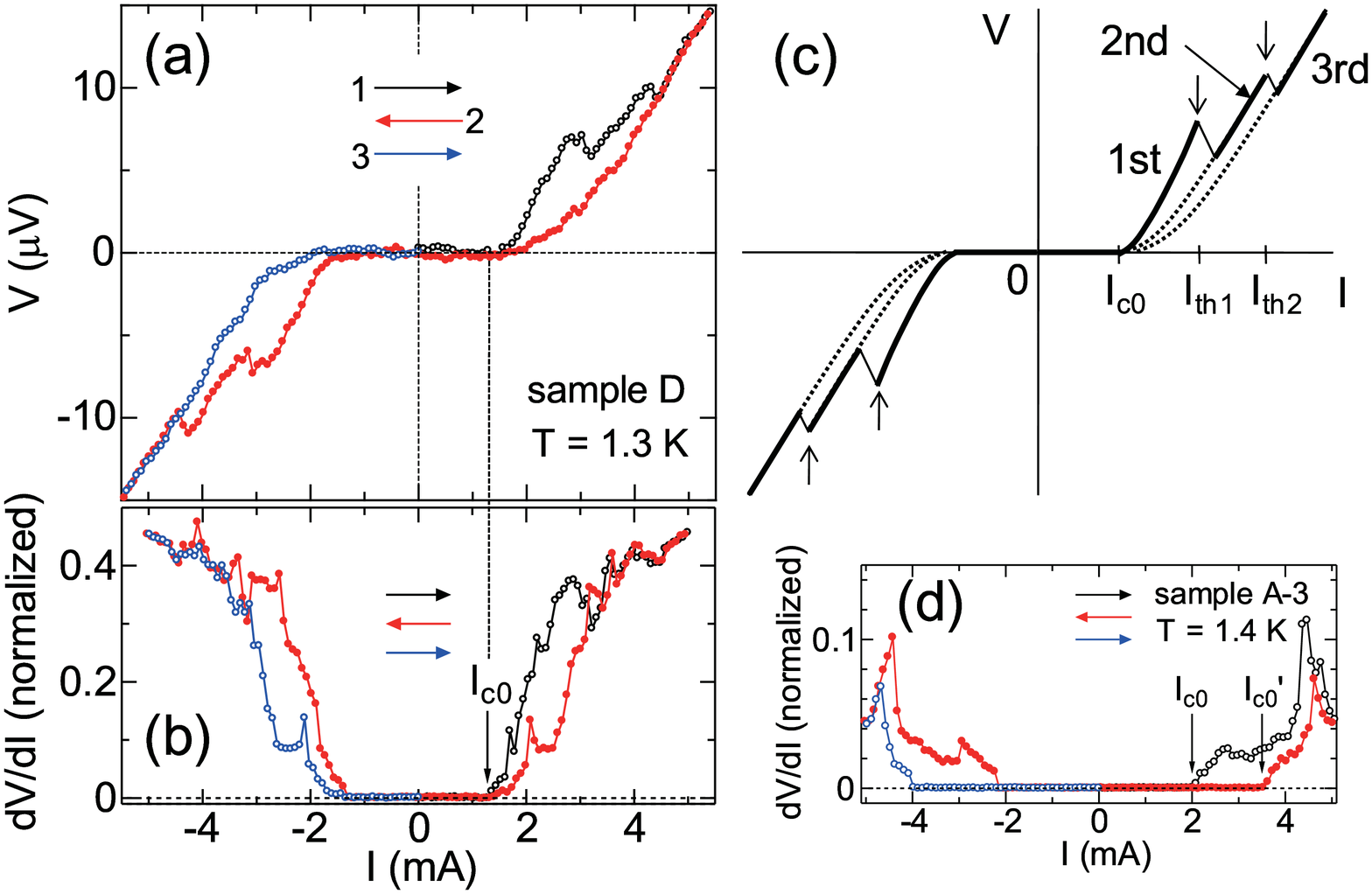}
\caption{\label{fig4}
(Color online)
(a) $V-I$ characteristics at 1.3 K obtained by DC method for sample D. The curves are obtained by averaging
over 10 curves. The open and filled symbols denote the different sweep directions, from zero to max.
(open black), max. to min. (filled red), and min. to zero (open blue).
(b) Normalized $dV/dI-I$ curves obtained by AC method for sample D.
(c) A schematic $V-I$ characteristics. At $I_{th1}$ ($I_{th2}$), the $V-I$ curve switches from
the 1st (2nd) to the 2nd (3rd) branch when DC current is swept up.
When DC current is swept down, the $V-I$ curve follows the 3rd branch.
(d) Normalized $dV/dI-I$ curves at 1.4 K for sample A-3. 
$I_{c0}$ and $I_{c0}'$ are critical currents when DC currents are swept up and down, respectively.}
\end{figure}

In the $dV/dI-I$ curves, we observed unusual hysteresis below 2 K,
as seen for sample A-3 (Fig. \ref{fig2}(b)).
The hysteresis curves are more pronounced at temperatures below $T_{c0}$
and tend to appear for small samples after FIB milling.
To demonstrate the anomalous behavior in $dV/dI-I$ clearly, we show $V-I$ curves obtained by DC method
as well as $dV/dI-I$ for sample D (the neck region is $3 \times 10 \times 5$ $\mu$m$^3$
(width$\times$length$\times$thickness)) at 1.3 K in Fig. \ref{fig4}(a)(b).
The $V-I$ data show discontinuous points at $\pm3$ mA and $\pm4.3$ mA when sweeping up of absolute
value of DC current (Fig. \ref{fig4}(a)).
To explain this behavior, we show a schematic of the $V-I$ in Fig. \ref{fig4}(c).
When we increase DC current from 0, a finite voltage appears at $I_{c0}$ like usual JJs.
Here the $I_{c0}$ is defined as a critical current at which a finite $dV/dI$ value is observed. 
Surprisingly, the voltage suddenly drops at a threshold $I_{th1} > I_{c0}$ denoted by the arrow.
Further increasing of DC current, a similar voltage drop appears at $I_{th2}$.
We note that one can measure only positive $dV/dI$ by AC method (Fig. \ref{fig4}(b)) because
the switching occurs instantaneously at $I_{th1}$, $I_{th2}$, etc. That is, $dV/dI$ value around
the threshold reflects that just before the switching or just after the switching.

Here we emphasize that $I_{th}$ is {\it not} a critical current from DC to AC Josephson effect
as seen in the inset of Fig. \ref{fig2}(b) because of the following anomalous features.
(i) At $I_{th}$, the voltage discontinuously {\it decreases} when the $V-I$ curve switches to the next branch.
(ii) It switches to a {\it lower}-$R_n$ (normal resistance) branch that has a larger $I_c$.
(iii) The hysteresis loop shows the opposite direction compared to typical JJs.
In serially connected typical JJs, e.g. the $c$-axis intrinsic JJs
of high-$T_c$ cuprate Bi$_2$Sr$_2$CaCu$_2$O$_{8+\delta}$ \cite{hkashiwaya},
the voltage absolutely increases and $R_n$ becomes larger after the zero-voltage state changes to
the finite-voltage state.
Furthermore in typical JJs, $I_c$ when sweeping up from 0 is {\it always}
higher than $I_{c0}'$ when sweeping down \cite{barone}, which is obviously contrary to the data
shown in Fig. \ref{fig4}(d).
Therefore, the anomalous switchings at $I_{th}$ above $I_c$ are never explained in terms of serially connected
JJs.
In other words, the anomalous switching phenomena occur in the {\it identical} channel.
Thus, it cannot be explained without considering the internal degrees of freedom of the superconducting state. 

\begin{figure}[ht]
\includegraphics[width=0.75\linewidth]{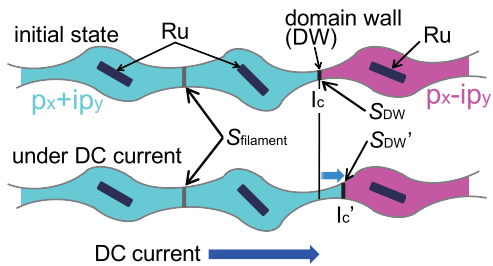}
\caption{\label{fig5}
(Color online)
Model of the chiral-domain-wall motion, induced by DC current.
$I_{c}$ and $I_{c}'$ are critical currents when DC currents are swept up and down, respectively.}
\end{figure}

One of the possible explanations of the unusual hysteresis is due to the chirality of the superconducting
state taking account of the 1.5-K phase being a chiral $p$-wave ($p_x \pm ip_y$) superconductor.
The chiral state has two types of distinct domains in the superconducting state.
If a superconducting linkage is formed from two antiparallel domains, there should exist
a domain wall (DW) between them. No DW is formed between parallel domains.
Let us assume that a local linkage channel contains both parallel and antiparallel domains,
as depicted in Fig. \ref{fig5}.
Each critical current is expected to be proportional to each cross section, i.e.
$I_c \propto S_{\rm filament}$ or $I_c \propto S_{\rm DW}$, where
$S_{\rm filament}$ is the cross section at which a weak link forms between parallel domains,
and $S_{\rm DW}$ is that between antiparallel domains.
Generally, a DW is likely to be formed and pinned at defects in the sample.
Once a DW forms, DC current which transfers a Cooper pair with a given chiral state
($p_x - ip_y$) to the opposite chiral state ($p_x + ip_y$), induces the DW motion as a back action.
Assuming the spatial variation of the cross section of the linkage channel, the movement of the DW
varies $I_c$ during the current sweep.
If the DW is pinned at defects at low-bias currents, the DW slides to the next metastable
position when the DC current beyond the threshold ($I_{th1}$, $I_{th2}$, etc. in Fig. \ref{fig4}(c))
is applied. Thus, the branch of $V-I$ switches, which causes the anomalous hysteresis.
As a small $S_{\rm DW}$ should be energetically favorable because the different order parameter
overlaps each other at the DW, it is reasonable that $I_{c0}$ is lower than $I_{c0}'$.
On the other hand, no hysteresis would appear in a channel between parallel domains without a DW.
In short, $dV/dI-I$ shows two-types: no hysteresis without a DW, and hysteresis with a DW moving
in alternate directions.
The DW motion is analogous to the current-driven DW motion in magnetic wires \cite{yamaguchi}. 
Here, we note that the chiral domain picture is consistent with the SM's prediction of the 2nd transition
at $T_2^{\ast}$ (3 K$> T_2^{\ast} >$1.5 K) with time reversal symmetry breaking.
We also note that no hysteresis has been reported \cite{hooper}, possibly because $I_c$ of averaged
channels smeared it under a large number of Ru inclusions.

Recent Kerr effect \cite{xia} and JJ \cite{kidwingira} experiments
suggest the presence of the chiral domains for the 1.5-K phase.
However, the estimated domain size is $\sim 50-100$ $\mu$m \cite{xia} and $\sim1$ $\mu$m \cite{kidwingira},
respectively. Thus, a consensus about the domain size is not established yet.
In our experiment for the 3-K phase, we estimate the domain size to be $\sim 10$ $\mu$m from the neck
region of the sample if the chiral domain scenario is correct.
We emphasize that our four-probe configuration is not sensitive to surface/interface states.
Thus, our result reflects the intrinsic properties of the superconducting state in SRO
removing experimental ambiguity as much as possible.
Further experimental work is needed to verify the presence of the chiral domains in the 3-K phase
and make further discussions about the effects of the chiral domains on the Josephson current \cite{asano}.

In summary, we revealed the superconducting nature of the 3-K phase in Sr$_2$RuO$_4$
by transport measurements on micro-fabricated samples.
We confirmed that the 3-K phase has odd-parity pairing symmetry, similar to the 1.5-K phase,
from the monotonous temperature dependence of the critical currents.
The unusual hysteresis of the differential resistance or the voltage-current characteristics
in the sweeping current observed below 2 K indicates the internal degrees of freedom of
the superconducting pairing, i.e. the chiral $p_x \pm ip_y$ state.
The domain wall motion induced by the DC current is a possible origin of the hysteresis.
This is a new discovery of the dynamic response of the superconducting order parameter of Sr$_2$RuO$_4$.

%
We are very grateful to T. Matsumoto for technical support.
This work was financially supported by a Grant-in-Aid for Scientific
Research on Priority Areas (No.~17071007) from MEXT, Japan.

%
%
%

%
%

\end{document}